\documentclass[letterpaper,twocolumn,fleqn]{article} 

\usepackage{ist}

\usepackage{amsmath} 
\usepackage{graphicx}
\usepackage{caption}
\usepackage{subfig}
\usepackage{lipsum}
\usepackage{multirow,booktabs}
\usepackage{multicol}
\usepackage{tabularx}

\usepackage{enumitem}
\usepackage{url}
\usepackage{hyperref}

\title{MS-UNIQUE: Multi-model and Sharpness-weighted Unsupervised Image Quality Estimation}

\author{ Mohit Prabhushankar\textsuperscript{1}, Dogancan Temel\textsuperscript{1}, Ghassan AlRegib\textsuperscript{1};\\
\textsuperscript{1}Center for Signal and Information Processing, School
	 of Electrical and Computer Engineering, Georgia Institute of Technology, Atlanta, GA}

\date{} 

\begin{document}

\onecolumn 

\begin{description}[labelindent=1cm,leftmargin=3cm,style=multiline]

\item[\textbf{Citation}]{M. Prabhushankar, D. Temel, and G. AlRegib, “MS-UNIQUE: Multi-model and Sharpness-weighted Unsupervised Image Quality Estimation”, the Electronic Imaging, Image Quality and System Performance XIV, Burlingame, California, USA, Jan. 29 – Feb. 2, 2017.} \\

\item[\textbf{DOI}]{\url{https://doi.org/10.2352/ISSN.2470-1173.2017.12.IQSP-223}} \\

\item[\textbf{Review}]{Publication date:  January 29, 2017} \\

\item[\textbf{Code/Slides}]{\url{https://ghassanalregib.com/publications/}} \\

\item[\textbf{Bib}] {
@article\{Temel2017\_EI,\\
title = "MS-UNIQUE: Multi-model and Sharpness-weighted Unsupervised Image Quality Estimation",\\
journal = "Electronic Imaging",\\
year = "2017",\\
number = "12",\\
publication date ="2017-01-29T00:00:00",\\
pages = "30-35",\\
itemtype = "ARTICLE",\\
issn = "2470-1173",\\
eissn = "2470-1173",\\
url = "https://www.ingentaconnect.com/content/ist/ei/2017/00002017/00000012/art00006",\\
doi = "doi:10.2352/ISSN.2470-1173.2017.12.IQSP-223",\\
author = "Prabhushankar, Mohit and Temel, Dogancan and AlRegib, Ghassan",\}\\
} \\


\item[\textbf{Contact}]{\href{mailto:alregib@gatech.edu}{alregib@gatech.edu}~~~~~~~\url{https://ghassanalregib.com/}\\ \href{mailto:dcantemel@gmail.com}{dcantemel@gmail.com}~~~~~~~\url{http://cantemel.com/}}
\end{description} 

\thispagestyle{empty}
\newpage
\clearpage

\twocolumn

\maketitle 

\thispagestyle{empty} 


\begin{abstract}
\vspace{+1.5mm}
In this paper, we train independent linear decoder models to estimate the perceived quality of images. More specifically, we calculate the responses of individual non-overlapping image patches to each of the decoders and scale these responses based on the sharpness characteristics of filter set. We use multiple linear decoders to capture different abstraction levels of the image patches. Training each model is carried out on 100,000 image patches from the ImageNet database in an unsupervised fashion. Color space selection and ZCA Whitening are performed over these patches to enhance the descriptiveness of the data. The proposed quality estimator is tested on the LIVE and the TID 2013 image quality assessment databases. Performance of the proposed method is compared against eleven other state of the art methods in terms of accuracy, consistency, linearity, and monotonic behavior. Based on experimental results, the proposed method is generally among the top performing quality estimators in all categories.

\end{abstract}

\section{Introduction}
\label{sec:intro}
\vspace{+1.5mm}
With the advent of social media and faster wireless networks, high quality digital images are one of the most popular forms of multimedia being shared online. Infact, on an average day, billions of photos are shared through dedicated platforms. It is essential for these platforms to maintain high standards in acquiring, compressing, transmitting, and displaying these images without compromising it's visual quality to the end user. Such a task cannot be manually performed due to it's mechanical and time consuming nature and the sheer volume of data involved. The goal of image quality assessment (IQA) is to automate this process by developing objective quality estimators that can predict subjective scores. In other words, the $perceived$ quality of images is measured objectively. Based on the availability of original distortion free images, image quality assessment algorithms are classified into three categories. Full-Reference (FR) metrics require the original image for predicting the quality of distorted image \cite{bib3,bib10,bib11,bib12,bib13,bib14,bib15,bib16}. No-Reference (NR) metrics estimate the quality of a distorted image without requiring access to the corresponding original image \cite{bib18,bib19,bib20}. Reduced-Reference metrics require a few feature sets extracted from the original image for quality prediction of the distorted image. In the proposed work, we focus on extending a FR model that we proposed in \cite{bib4} which was based on a data driven approach. 

Data driven approaches are not uncommon in IQA literature. The authors in \cite{bib17} propose MLIQM, a metric that benefits from the already present IQA theory to construct features, and apply SVM classification to understand the quality class. Then a SVM regression is used to estimate the quality of a distorted image within that quality class. The authors in \cite{bib18} apply a pre-training step in which they distort high quality images and feed them into their deep network to train a model that predicts the subjective score. The authors in \cite{bib19} propose an image quality assessment approach based on learning a set of filters through Support Vector Regression. The weights of SVR are learnt through a Stochastic gradient descent algorithm and their responses are used to estimate quality. In \cite{bib20}, the authors propose  an  unsupervised  learning approach to obtain quality-aware filters using distorted images. These filters are used to extract features that are then regressed using a random forest to obtain quality estimates. However, the common thread in all these algorithms, is the requirement of distorted images and subjective scores during training.

In this paper, we explore the combination of unsupervised learning and hand-crafting to extend learning networks to assess quality of images. In \cite{bib4} we had proposed UNIQUE, a shallow learning architecture to estimate quality. It had one hidden layer which was trained using a sparsity criterion where the weights and bias were considered a domain transformation on non-overlapping patches of images. This technique outperforms majority of the existing methods in LIVE \cite{bib8} and TID $2013$ \cite{bib9} databases. It is an unsupervised architecture since it does not require any target labels during the neural network training. Also, there is no need for either subjective scores or distorted images during training. Keeping all these advantages intact, we extend UNIQUE and improve it's performance by analyzing the weights which we learnt, utilizing existing IQA literature that stresses the importance of sharpness in measuring quality \cite{bib7}. We also learn multiple self-contained and reversible representations of undistorted data and use these representations to estimate quality of images. We propose MS-UNIQUE which is a full reference image quality assessment algorithm based on an unsupervised learning approach through distortion-free images. 

\section{Methodology}
\label{TrainingAndMethodology}
\vspace{+1.5mm}
We propose learning a set of weights and bias from a linear decoder. Before using the learning framework, we preprocess the data to make it more descriptive. The learnt weights from a linear decoder are considered as a filter set which are used to estimate the quality of an image. And if linear decoder models with different number of neurons in the hidden layer are trained, we obtain a number of filter sets all learning to model the same input using multiple representations.  The filters are also made structure aware by differentiating the ones that capture edges from the ones that capture color.

\subsection{Color Space Selection}
\vspace{+1.5mm}
We use luminance (Y channel) as part of our input data. The human visual system is more sensitive towards changes in intensity domain rather than chroma \cite{bib1}. The authors in \cite{bib3} claim that structural information can be gleaned from the normalized luma domain. In addition to the Y channel, we use the green channel from RGB color space. Green channel is selected since it contains a large part of the information from R and B color channels. This is verified by measuring the cross correlation between channels of RGB representations - the cross correlation $r_{RG}$ between R and G color channels is $0.98$ and $r_{GB}$, between G and B color channels is $0.94$ \cite{bib2}. We augment the Y and the G channels with the Cr channel after a transformation into YCbCr colorspace. This is done to include chroma information as part of our data. The specific plane Cr is chosen over Cb based on experimental results. The three planes are combined to obtain a descriptive YGCr image.

\subsection{Data Matrix Preparation}
\begin{figure}[htbp!]
	\begin{center}
		\noindent
		\includegraphics[width=0.9\linewidth]{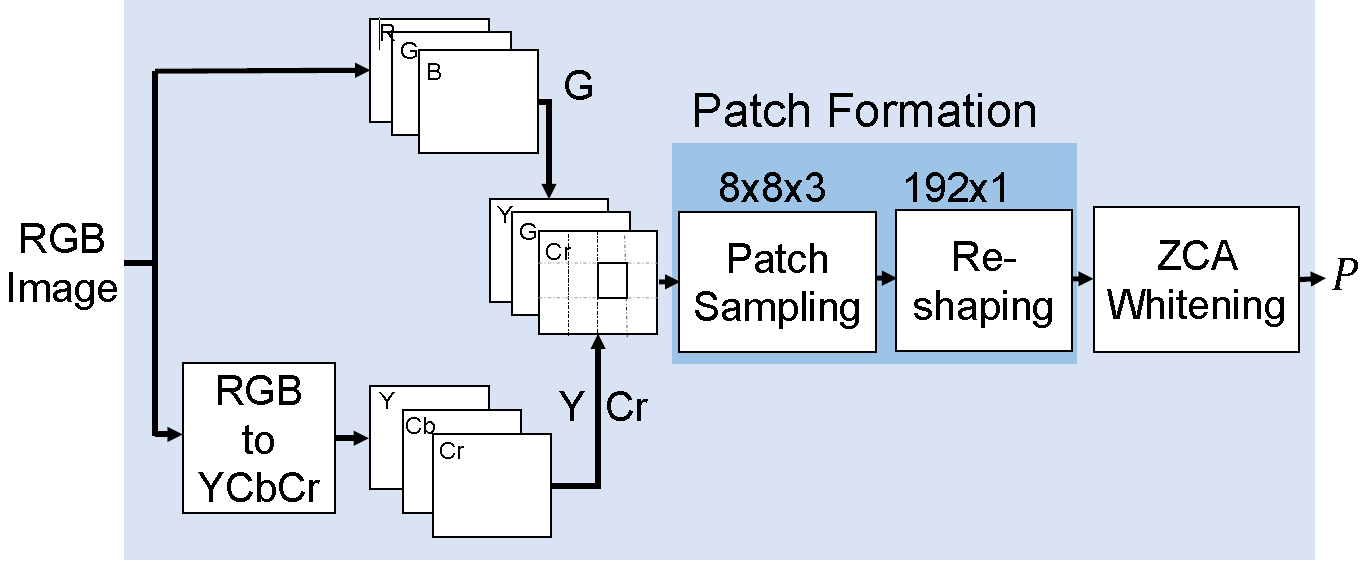}
		\vspace{1 mm}
		\caption{Data matrix preparation.}
		\label{fig:pre}
	\end{center}
\end{figure}

\vspace{+1.5mm}
From the ImageNet 2013 test database, $1,000$ images are randomly selected during training. We do not use any annotated metadata associated with the images. Each image is first transformed into YGCr colorspace. From each image, we extract $100$ patches of size $8$x$8$x$3$ randomly. Each patch is then reshaped into a $192$x$1$ column vector. The patch vectors from all images are stacked together to get a $192$x$100000$ input patch matrix. The data matrix is then passed through a Zero Component Analysis (ZCA) Whitening algorithm. Whitening is performed to decorrelate adjacent pixels in raw data so as to lessen redundancy. The authors in \cite{bib6} show that the HVS performs whitening. Essentially, this converts the input data with a zero mean covariance matrix into whitened data with an identity covariance matrix. The adjacent features in the input matrix are decorrelated and the variance of each is one. ZCA also satisfies the property that the whitening matrix is orthogonal. Note that whitening is not performed on the $100,000$ patches but on the $192$ input features in each patch feature vector. Hence, individual pixels inside a patch are decorrelated from other pixels in the same patch. This happens over all $100,000$ patches hence lowering the redundancy fed into the learning architecture from each feature vector \cite{bib5}. We summarize the data matrix preparation in Figure \ref{fig:pre}.
\subsection{Linear Decoder}
\begin{figure*}%
	\centering
	\subfloat[81 neurons]{{\fbox{\includegraphics[height = 0.2\linewidth,width=0.19\linewidth]{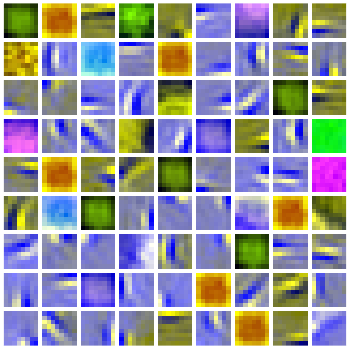} }}}%
	\subfloat[121 neurons]{{\fbox{\includegraphics[height = 0.2\linewidth,width=0.19\linewidth]{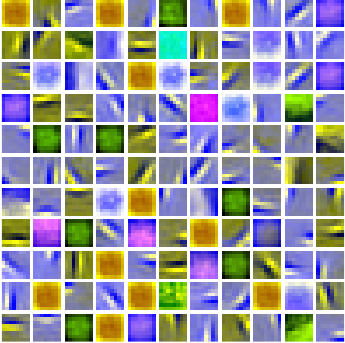} }}}%
	\subfloat[169 neurons]{{\fbox{\includegraphics[height = 0.2\linewidth,width=0.19\linewidth]{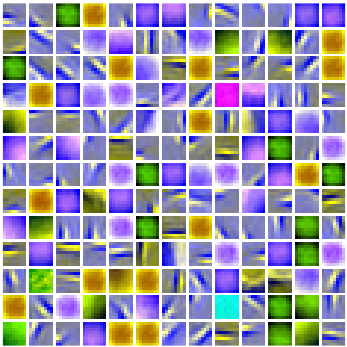} }}}%
	\subfloat[400 neurons]{{\fbox{\includegraphics[height = 0.2\linewidth,width=0.19\linewidth]{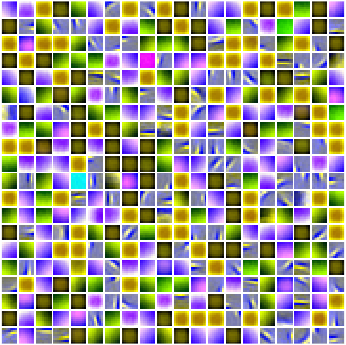} }}}%
	\subfloat[625 neurons]{{\fbox{\includegraphics[height = 0.2\linewidth,width=0.19\linewidth]{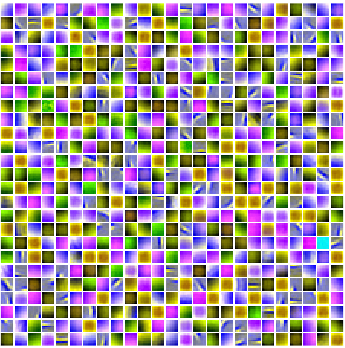} }}}%
	\caption{Weight Visualizations. In each set, each square can be used to infer input patches that maximally activate it. Each individual square in all sets is of size $8$x$8$x$3$ and is scaled here for visualization purposes.}%
	\label{fig:Visualization}%
\end{figure*}

\vspace{+1.5mm}
A linear decoder is an unsupervised neural network framework used to represent data in different dimensions. In this work, we use a framework with only one hidden layer. It can be used to sparsify data or learn a compact representation by changing the number of neurons in the hidden layer. The framework functions by transforming the input into hidden layer activations or responses and then reconstructing back the input using these responses. Transformation occurs through a set of weights and bias that are randomly initialized and then adjusted iteratively based on the reconstruction error using backpropagation. The hidden layer responses are obtained as
\begin{equation}\label{ForwardFilters}
s = sigmoid(W_1^T P + b_1),
\vspace{-1.5mm}
\end{equation} 
where $s$ is the response, $W_1$ and $b_1$ are the forward weights and bias. Each column in $W_1$ is a $192$x$1$ vector that filters each patch from the data. Sigmoid is the non-linear layer used in our framework. These hidden layer responses are filtered using another set of backward weights $W_2$ and bias $b_2$ to obtain back a reconstructed version of the input $\tilde{P}$ as

\begin{equation}\label{BackwardFilters}
\tilde{P} = W_2^T s + b_2,
\vspace{-1.5mm}
\end{equation} 
Note that there is no sigmoid layer after reconstruction. The objective function for backpropagation $J(W_1,W_2,b_1,b_2)$ is given by 
\begin{equation}
	J(W,b) = \lVert (W_2^Ts + b_2) - P \rVert_2^2 + \beta \sum_{j=1}^{N} {\rm KL}(\rho || \hat\rho_j) + \lambda\lVert W \rVert_2^2,
\end{equation}
where the first term is the reconstructed $L2$ norm error, the second term is the sparsity penalty term and the third is the weight decay or regularization term. $N$ is the total number of patches, which amounts to $100,000$. Sparsity penalty is included to constrain the average activation of neurons to be close to zero and this penalty is obtained using KL-Divergence over all training patches. $\rho$ is the desired average activation which is set to $0.035$. The sparse penalty term goes to $0$ when the actual average activation $\hat{\rho}$ comes close to $\rho$. It is weighted by $\beta$ which is set to $5$. The weight decay term $\lambda$, which is set to $3e^{-3}$ acts as a regularization term by decreasing the magnitude of weights thereby preventing overfitting.  
\begin{figure}[htbp!]
	\begin{center}
		\noindent
		\includegraphics[width=0.6\linewidth]{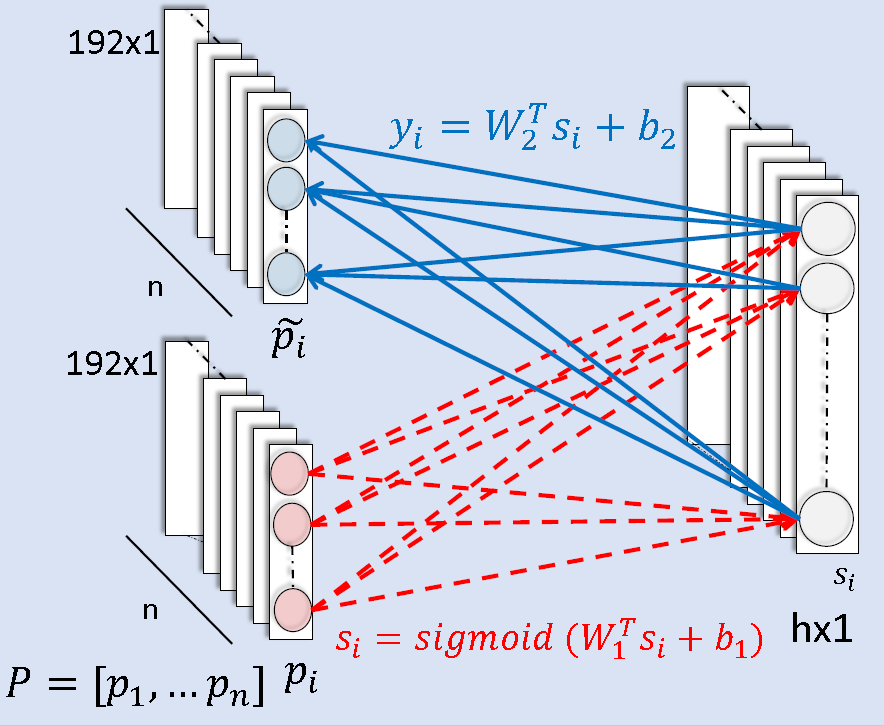}
		\vspace{+1 mm}
		\caption{Linear decoder architecture}
		\label{fig:decoder}
	\end{center}
\end{figure}
We show the architecture of a linear decoder in Figure \ref{fig:decoder} in which $h$ corresponds to the number of neurons in the hidden layers. We change the number of neurons $h$ to obtain models that can either sparsely or compactly represent the input data. We visualize the weight sets corresponding to different values of $h$ in Figure \ref{fig:Visualization}.

\begin{figure}%
	\centering
	\subfloat[Edge Filters]{{\fbox{\includegraphics[height = 0.45\linewidth,width=0.45\linewidth]{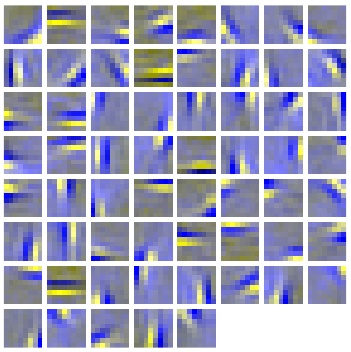} }}}%
	\subfloat[Color Filters]{{\fbox{\includegraphics[height = 0.45\linewidth,width=0.45\linewidth]{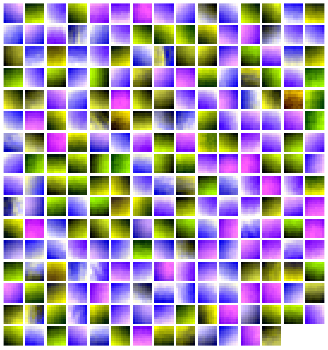} }}}%
	\caption{Result Visualization of differentiating a 625 filter model into edge and color aware filters.}%
	\label{fig:Edge/Color}%
\end{figure}
\subsection{Multi-model training}
\vspace{+1.5mm}
The data matrix is fed into a linear decoder model with $h$ = $81$ and trained for $400$ epochs. The trained forward weights and bias are stored. This step is repeated to obtain weights and bias for $h$ = $121,169,400,625$ separately. The sparsity parameter during training ensures that none of the filters from any model get activated abnormally over the others. This multi-model approach ensures that we represent an image patch both sparsely and compactly and learn multiple filter sets that combine non-linearly to reconstruct it. Also, a sparse filter set learns more localized features while a compact set learns global features.

\subsection{Sharpness aware filters}
\vspace{+1.5mm}
Sharpness is an important determining factor in the perceptual quality assessment of images \cite{bib7}. The HVS is adept at detecting blur and evaluating quality based on sharpness. However, our learning framework does not use any handcrafted features like incorporating edges. Hence we add this feature to the already constructed filter set. We make our filters sharpness aware by analyzing their descriptiveness and then weighing their responses accordingly. We give higher importance to filters that capture edges rather than color. Distinguishing filters based on edge characteristics is performed using the bias corrected implementation of kurtosis. Kurtosis is defined as,
\begin{equation}\label{Kurtosis}
k = \frac{E(x-\mu)^{4}}{\sigma^4}
\vspace{-5mm}
\end{equation} 
Hence, further away a data point is from the mean of the distribution, larger is it's influence on kurtosis. We theorize that filters that capture edge components consist of more data points that are away from the mean of the overall data making them outliers. The presence of these outliers gives a higher kurtosis score to edge filters. The kurtosis of each vectorized, zero centered, and normalized filter is measured against a threshold to capture it's edge characteristics. Any filter with a kurtosis greater than $5$ is labeled as an edge filter while filters with kurtosis less than $2$ are labeled as color filters. The results of thresholding on a $625$ filter model set is shown in Figure \ref{fig:Edge/Color}. 

\begin{figure}[htbp!]
	\begin{center}
		\noindent
		\includegraphics[width=1\linewidth]{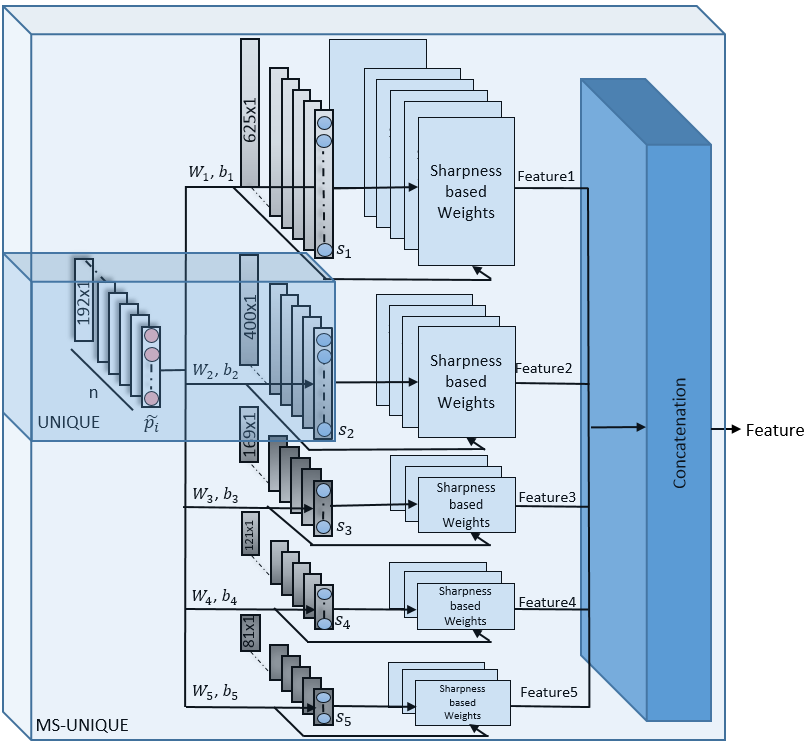}
		\vspace{+1 mm}
		\caption{Feature generation}
		\vspace{-5.0mm}
		\label{fig:FeatureGeneration}
	\end{center}
\end{figure}
\begin{table*}[!htb]
	\scriptsize
	\centering
	\caption{Performance of image quality estimators.}
	\label{tab_results_databases}
	\begin{tabular}{|l|l|l|l|l|l|l|l|l|l|l|l|l|}
		\hline
		
		\multirow{3}{*}{{\bf Methods}}                & \multirow{2}{*}{{\bf PSNR}}  & {\bf PSNR } &{\bf PSNR } &\multirow{2}{*}{{\bf SSIM }} &{\bf MS } &{\bf CW }&{\bf IW } &{\bf SR }   &\multirow{2}{*}{{\bf FSIMc }} &\multirow{2}{*}{{\bf PerSIM }} &\multirow{2}{*}{{\bf UNIQUE }}  &{\bf MS- } \\
		& &\textbf{HA} &\textbf{HMA} & &{\bf SSIM} &{\bf SSIM} &{\bf SSIM } &{\bf SIM }  &&&&{\bf UNIQUE} 
		\\ 
		& &\cite{bib10} & \cite{bib10} & \cite{bib3} & \cite{bib11} &\cite{bib12} & \cite{bib13} &\cite{bib14}  & \cite{bib15} & \cite{bib16} & \cite{bib4} & \\
		\hline 
		
		& \multicolumn{12}{c|}{\textbf{Outlier Ratio}}                                                                                                                                                                        
		
		
		\\ \hline
		\textbf{TID13}   
		&0.725   & 0.615   & 0.670  &  0.732  &  0.697 &    0.855 &    0.700 &   0.632 &    0.727 &    0.655  &0.640 & \bf 0.611
		
		\\ \hline

		& \multicolumn{12}{c|}{\textbf{Root Mean Square Error}}                                                                                                                                                                        \\ \hline
		\textbf{LIVE}  &8.61   & 6.93   & \bf 6.58  &  7.52  &  7.43 &  11.2 &   7.11 &   7.54 &   7.20 &   6.80  & 6.76  &  6.61 \\ 
		\textbf{TID13}     &0.87   & 0.65  &  0.69 &   0.76 &   0.68  &  1.20  &  0.68 &  0.61  &  0.68 &   0.64 & 0.60  & \bf 0.57
		\\ \hline              
		
		& \multicolumn{12}{c|}{\textbf{Pearson Correlation Coefficient}}                                                                                                                                                                        \\ \hline
		
		\multirow{1}{*}{{\bf LIVE}}                 & 0.928             &0.953     & \bf 0.958             & 0.945                  &0.946           &0.872                   &0.951                  & 0.945               &0.950                                &0.955             & 0.956      & \bf 0.958                   \\


		\multirow{1}{*}{{\bf TID13}}       &0.705                & 0.850                  &0.827                    &0.789         &0.832                 &0.227                  &0.831                 & 0.866            &0.832                              & 0.854       & 0.870 & \bf 0.884     \\ \hline                         

		\textbf{}      & \multicolumn{12}{c|}{\textbf{Spearman Correlation Coefficient}}                                                                                                                                                                        \\ \hline

		\multirow{1}{*}{{\bf LIVE}}             &0.909                &0.937                 &0.944                 &              0.949  & 0.951                 & 0.902                  & \bf 0.960                 & 0.955                        & 0.959              & 0.950      &   0.952    & 0.949 \\ 


		\multirow{1}{*}{{\bf TID13}}          & 0.700              & 0.847                 & 0.817                  &0.741        &0.785                & 0.562               & 0.777              &0.807             &0.851                             & 0.853      &0.860  &\bf 0.870  
		\\ \hline

	\end{tabular}
\end{table*}


\begin{table*}[!htbp]
	\centering
	\scriptsize
	\caption{Distributional difference between subjective scores and objective quality estimates.}

	\begin{tabular}{|c||c|c|c|c|c||c|c|c|c|c|}   \hline
		\multirow{2}[4]{*}{\textbf{Metric}} 
		
		& \multicolumn{5}{c||}{\textbf{Difference-LIVE}}  & \multicolumn{5}{c|}{\textbf{Difference-TID13}} \\ \cline{2-11}     
		
		& \textbf{EMD}& \textbf{KL}& \textbf{JS}& \textbf{HI}& \textbf{L2}&  \textbf{EMD}& \textbf{KL}& \textbf{JS}& \textbf{HI}& \textbf{L2}    
		\\ \hline 
		
		\textbf{PSNR-HMA}  &  0.226  &  0.205  &  0.053  &  0.226   & 0.066 &     0.360  &  0.927 &   0.117  &  0.360 &   0.124
		\\ \hline
		
		\textbf{IW-SSIM}  &  0.297  &  0.325  &  0.072 &   0.297  &  0.076  &   0.500 &   1.678 &   0.196 &   0.500  &  0.180
		\\ \hline

		\textbf{UNIQUE}  &   0.236  &  0.258 &   0.055  &    0.236  &  0.069 &     0.386 &   0.855 &   0.120 &   0.386  & 0.109
		\\ \hline
		
		\textbf{MS-UNIQUE}  &  \bf 0.209  & \bf 0.176 &  \bf 0.038  &   \bf 0.209  & \bf 0.057 &    \bf  0.357 &  \bf 0.734 & \bf  0.108 &  \bf 0.357  & \bf 0.103
		
		\\ \hline
	\end{tabular}%
	\label{tab:hist_dist}
	\vspace{-4.0mm}
	
\end{table*} 

\subsection{Image Quality Assessment}
\vspace{+1.5mm}
We preprocess images as described previously and utilize the formulation in Eq.\ref{ForwardFilters} to obtain filter responses. These responses are weighted based on the sharpness characteristics of corresponding filters. The edge filter responses are given a higher weightage of $2$ while the color responses are lowered by a weight of $0.5$. This is performed for all models to obtain one feature vector per image. The feature generation process is summarized in Figure \ref{fig:FeatureGeneration}. The responses in feature vector that are significantly less than the average activation value set during training are assigned a zero to mimic the suppression mechanisms in the HVS. We generate feature vectors for both reference and distorted images. The feature vectors corresponding to the original and distorted images are compared using $10^{th}$ power of Spearman correlation coefficient to fully utilize quality estimation range.    

The proposed method is an extension of the quality estimator UNIQUE \cite{bib4} as shown in Figure \ref{fig:FeatureGeneration}. It builds on UNIQUE by weighing filter responses. We also propose using multiple decoders with different number of neurons in the hidden layer to abstract local and global characteristics in image patches.

\section{Validation}
\vspace{+1.5mm}
\subsection{Database}
\vspace{+1.5mm}
The proposed quality estimator is validated on the LIVE image quality \cite{bib8} and TID 2013 \cite{bib9} databases. The databases have more than $3500$ distorted images between them. These images can be classified into $7$ categories based on their distortion types - compression artifacts, image noise, communication errors, blur artifacts, color degradations, global, and local distortions. The compression artifacts category consists of the JPEG and the JPEG2000 compressions, and lossy compressions of noisy images. The noise category includes Gaussian noise and additive noise added in color components, spatially correlated noise, masked noise, high frequency noise, impulse noise, quantization noise, image denoising, multiplicative Gaussian noise, and comfort noise. The communication errors category includes the JPEG and the JPEG2000 transmission errors of noisy images. The blur artifacts category consists of Gaussian blur, and sparse sampling and reconstruction. The color degradations category contain changes in color saturation, image color quantization with dither, and chromatic aberrations. The global category includes intensity shifts, and contrast changes while the local category contains non-eccentricity pattern noise, and local blockwise distortions of different intensities.

\subsection{Performance Metrics}
\vspace{+1.5mm}
Validation of MS-UNIQUE and compared algorithms are carried out in terms of root mean square error, outlier ratio, Pearson and Spearman correlation coefficients. In the outlier ratio calculations, we use those data points that lie two standard deviations away from the average subjective scores. Also, outlier ratio is only reported for TID 2013 database since the standard deviations of subjective scores are not reported in LIVE database. The regression formulation from \cite{bib8} is used to calculate regress estimates of all methods before comparing. We report the difference between the normalized histograms of subjective scores and the regressed quality estimates through common histogram difference metrics including Earth Mover’s Distance (EMD), Kullback-Leibler (KL) divergence, Jensen-Shannon (JS) divergence, histogram intersection (HI), and L2 norm.

\subsection{Results}
\vspace{+1.5mm}
The proposed quality estimator is compared against eleven other commonly used or state of the art full reference quality assessment methods based on fidelity, perceptually-extended fidelity, structural similarity, feature similarity, and perceptual similarity. The performances of all these metrics are summarized in Table \ref{tab_results_databases} with the highest performing metric in each category displayed in bold. PSNR-HMA, IW-SSIM, UNIQUE, and MS-UNIQUE are among the top performing metrics. MS-UNIQUE outperforms all these estimators in TID13 database among all performance metrics. In the LIVE database it consistently performs well in all but two of the metrics. IW-SSIM performs better in terms of SROCC in this database. However, MS-UNIQUE outperforms IW-SSIM among all the other categories. Both MS-UNIQUE and PSNR-HMA provide similar results in terms of PCC. MS-UNIQUE's results for RMSE are slightly lesser than PSNR-HMA. MS-UNIQUE builds on UNIQUE among all categories except in SROCC in LIVE database.

\begin{figure}[!htb]%
	\centering
	\subfloat[LIVE PSNR-HMA]{{\includegraphics[height = 0.4\linewidth,width=0.5\linewidth]{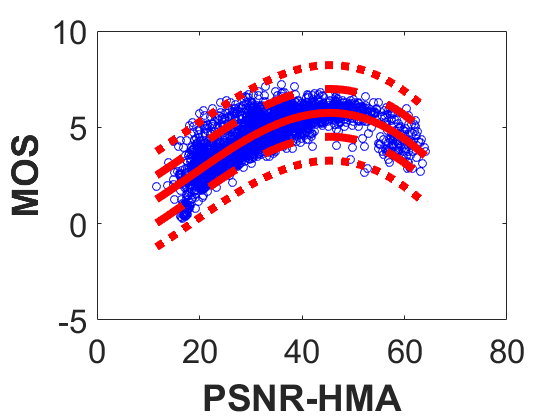} }}%
	\subfloat[TID PSNR-HMA]{{\includegraphics[height = 0.4\linewidth,width=0.5\linewidth]{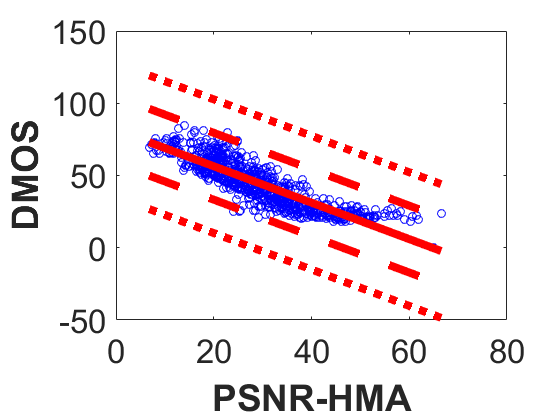} }}%
	\qquad
	\subfloat[LIVE IW-SSIM]{{\includegraphics[height = 0.4\linewidth,width=0.5\linewidth]{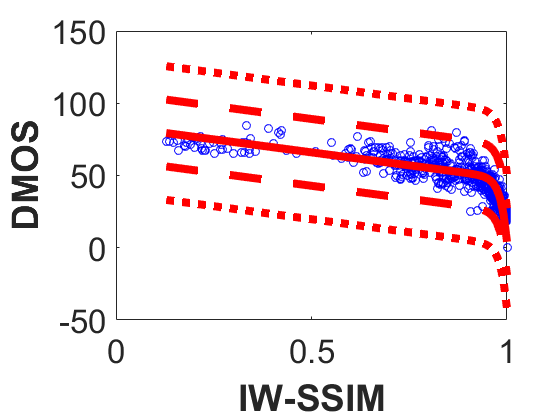} }}%
	\subfloat[TID IW-SSIM]{{\includegraphics[height = 0.4\linewidth,width=0.5\linewidth]{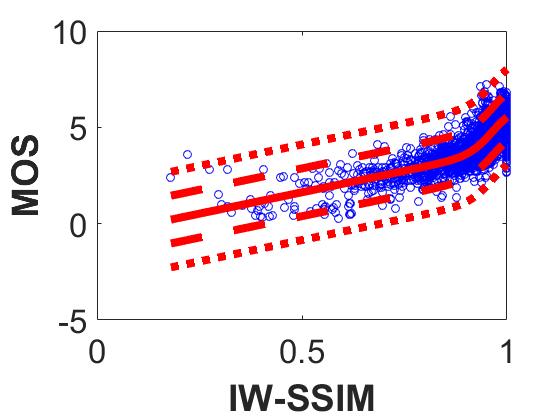} }}%
	\qquad
	\vspace{-3 mm}
	\subfloat[LIVE UNIQUE]{{\includegraphics[height = 0.4\linewidth,width=0.5\linewidth]{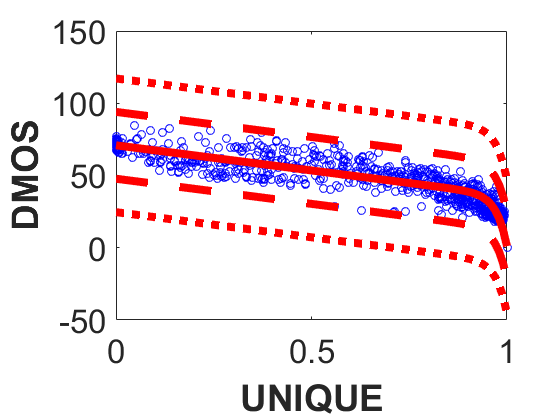} }}%
	\subfloat[TID UNIQUE]{{\includegraphics[height = 0.4\linewidth,width=0.5\linewidth]{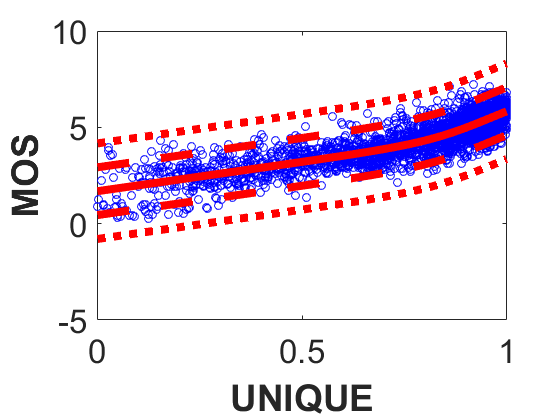} }}%
	\qquad
	\vspace{-3 mm}
	\subfloat[LIVE MS-UNIQUE]{{\includegraphics[height = 0.4\linewidth,width=0.5\linewidth]{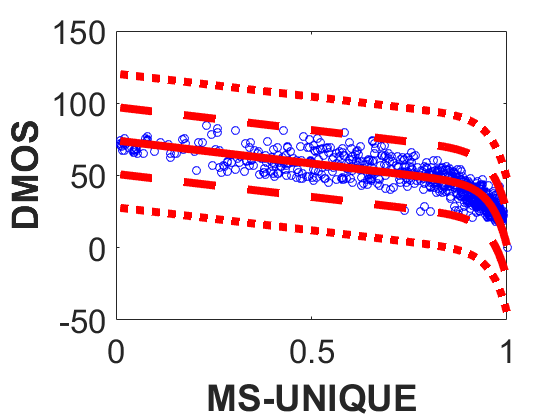} }}%
	\subfloat[TID MS-UNIQUE]{{\includegraphics[height = 0.4\linewidth,width=0.5\linewidth]{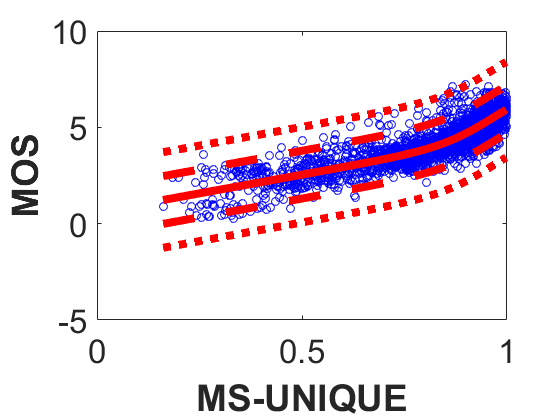} }}%
	\vspace{+5 mm}
	\caption{Scatter plots of top performing quality estimators}%
	\label{fig:ScatterPlots}%
\end{figure}

To better analyze the distribution of subjective scores against the estimated scores, scatter plots of the best performing metrics are shown in Figure \ref{fig:ScatterPlots}. The X-axis corresponds to the estimated scores while the Y-axis is the ground truth subjective mean opinion scores (MOS) or differential Mean Opinion Scores (DMOS). For an ideal quality estimator, the scatter plot data should follow a linear curve with low deviation. This is not observed in PSNR-HMA which shows a parabolic curve in LIVE database. There is a much sharper drop off in IW-SSIM with most of the points concentrated on the far end of the curve in LIVE database. UNIQUE and MS-UNIQUE have a far more linear curve with scores extending throughout the range. To numerically differentiate between MS-UNIQUE and other metrics in terms of regressed quality estimates, we present the difference between the normalized histograms of ground truths and regressed results, in Table \ref{tab:hist_dist}. The best results are highlighted in bold and MS-UNIQUE consistently performs well in both the databases among all compared metrics. Overall, MS-UNIQUE is the best performing metric in $15$ out of $17$ compared metrics over both databases.

\section{Conclusion}

We proposed an extension to the quality estimator UNIQUE, by analyzing the learning network used and handcrafting a weighing scheme that captures sharpness. This is done in the preprocessing and postprocessing blocks by enhancing information acquired from the data, analyzing the edge characteristics of learnt filters so that their responses are weighed based on quality assessment theory. Multiple models of linear decoders, where the number of hidden layer neurons represent the local or global characteristics captured, are used to obtain different levels of abstraction. The performance of MS-UNIQUE shows that performance of metrics that use a data driven approach can be enhanced by handcrafting features.



\begin{biography}
\vspace{+1.5mm}
Mohit Prabhushankar received the M.S. degree in Electrical and Computer Engineering with a minor in Computer Science from Georgia Institute of Technology, Atlanta, in 2015. Since then, he has been pursuing Ph.D. degree in the Center for Signal and Information Processing (CSIP), Georgia Institute of Technology, USA, as a Research Assistant. His research interests include image quality assessment, image denoising and enhancement feature design through data driven approaches.

Dogancan Temel received an M.S. degree with a minor in Management in 2013, and a PhD degree
with a minor in Computer Science in 2016 from the school of Electrical and Computer Engineering
in Georgia Institute of Technology, Atlanta. While his studies at Georgia Tech, Dr. Temel worked in
the Multimedia and Sensors Lab at the Center for Signal and Information Processing as a Graduate
Research Assistant and in Texas Instruments as a Systems Engineering intern. Dr. Temel worked on
various projects including perceived image quality assessment, deep learning-based image processing
and computer vision, high color range imaging, vital sign monitoring, computational aesthetics, seis-
mic interpretation, 3D reconstruction, streaming, and quality assessment. 

Ghassan AlRegib is currently Associate Professor at the School of Electrical and Computer Engineering at the Georgia
Institute of Technology in Atlanta, GA, USA. His research group is working on projects related to image and video
processing and communications, immersive communications, collaborative systems, quality of images and videos, and
3D video processing. Prof. AlRegib is a Senior Member of IEEE. Prof. AlRegib served as the chair of the Special Sessions Program at the IEEE International Conference
on Image Processing (ICIP)  He was the Track
Chair in the IEEE International Conference on Multimedia and Expo (ICME) in 2011 and the co-chair of the IEEE MMTC Interest Group on
3D Rendering, Processing, and Communications, 2010-present. Prof. AlRegib is a member of the Editorial Board of the Wireless Networks
Journal (WiNET), 2009-present. Prof. AlRegib co-founded the ICST International Conference on Immersive Communications (IMMERSCOM)
and served as the Chair of the first event in 2007. Prof.
AlRegib is the founding Editor-in-Chief (EiC) of the ICST Transactions on Immersive Communications to be inaugurated in late 2012. He is
also the Chair of the Speech and Video Processing Track at Asilomar 2012.

\end{biography}

\end{document}